\documentclass[preprint,english,showpacs,11pt,floatfix]{revtex4}

\usepackage{babel,amsmath,amssymb,dcolumn}
\usepackage[dvips]{graphics}

\begin{document}

\title{Dirac Cosmology and the Acceleration of the Contemporary Universe}

\author{Cheng-Gang Shao, Jianyong Shen, Bin Wang}
\email{wangb@fudan.edu.cn} \affiliation{Department of Physics,
Fudan University, Shanghai 200433, People's Republic of China }

\author{Ru-Keng Su}
\email{rksu@fudan.ac.cn}\affiliation{China Center of Advanced
Science and Technology (World Laboratory) P.O. Box 8730, Beijing
100080, People's Republic of China}\affiliation{Department of
Physics, Fudan University, Shanghai 200433, People's Republic of
China }

\begin{abstract}
A model is suggested to unify the Einstein GR and Dirac Cosmology.
There is one adjusted parameter $b_2$ in our model. After
adjusting the parameter $b_2$ in the model by using the supernova
data, we have calculated the gravitational constant $\bar G$ and
the physical quantities of $a(t)$, $q(t)$ and $\rho_r(t)/
\rho_b(t)$ by using the present day quantities as the initial
conditions and found that the equation of state parameter
$w_{\theta}$ equals to -0.83 , the ratio of the density of the
addition creation $\Omega_{\Lambda}=0.8$ and the ratio of the
density of the matter including multiplication creation, radiation
and normal matter $\Omega_m =0.2$ at present. The results are
self-consistent and in good agreement with present knowledge in
cosmology. These results suggest that the addition creation and
multiplication creation in Dirac cosmology play the role of the
dark energy and dark matter.
\end{abstract}


\maketitle

\section{Introduction}
According to the Dirac's arguments, the large dimensionless
numbers provided by atomic physics and astronomy of our universe
are connected with each other \cite{s1}. These numbers include:
(i) the ratio of the electric to the gravitational force between
an electron and a proton $ a_1  = {\raise0.7ex\hbox{${e^2 }$}
\!\mathord{\left/
 {\vphantom {{e^2 } {Gm_p m_e }}}\right.\kern-\nulldelimiterspace}
\!\lower0.7ex\hbox{${Gm_p m_e }$}} \sim 10^{39}$; (ii) the age of
the universe expressed in terms of the atomic unit $ a_2  =
\frac{{m_e c^3 }}{{e^2 H}} \sim 10^{39}$, where $H$ is the Hubble
constant; (iii) the mass of the part of the universe which is
receding from with a velocity $v<c/2$, expressed in the units of the
proton mass, say, $ a_3  \sim 10^{78}$. Dirac introduced a large
number hypothesis
\begin{equation} \label{e1}
a_1  \cong a_2  \cong a_3 ^{1/2}
\end{equation}
Based on this hypothesis, Dirac suggested a model of cosmology with
a varying gravitational constant $G$ and an increase in the amount
of matter in the universe.

Based on the large number hypothesis, a number of cosmological
models with a varying gravitational constant have been proposed
\cite{s2}-\cite{s10}. However most of them met many difficulties.
Noting that gravitational constant $G$ cannot vary in general
relativity (GR), the first difficulty is that one must explain the
contradiction between Einstein GR and Dirac Cosmology. Though much
effort, for example, Milne two time scale hypothesis \cite{s6},
Weyl's geometry \cite{s1} etc., has been devoted to reconcile the
requirements of these two theories, it is still an open question to
establish a theory which can unify the Dirac cosmology and Einstein
GR.

The second difficulty comes from experiments. Almost all experiments
at the scales of solar system and galaxies have not found the
variation of $G$ \cite{s11}-\cite{s12}. A possible variation of $G$
has been investigated with no success through geophysical and
astronomical observations. From experimental results one tends to
believe that $G$ is a constant for local system with large scale.

The third difficulty belongs to the conservation of energy and
momentum. Usually we use a perfect-fluid energy-momentum tensor to
describe the matter of universe and it is conserved in the cosmic
evolution. The addition creation, the new matter created uniformly
in the whole space, and the multiplication creation, the new
matter created in regions where old matters exist, must come from
other mechanism as suggested by Dirac.

Many years ago a possible unified theory of Dirac cosmology and GR
to overcome the above difficulties was suggested in \cite{s2}. The
basic idea is as follows. Though one would expect a constant value
of $G$ at the local system such as solar system, binary system and
galaxies, it must be stressed that the cosmological observations
still cannot put strong limits on the time variation of $G$ in the
cosmological scale, especially at the late time of the universe
evolution \cite{s10}\cite{s11}. Though Einstein GR has been proved
to be correct by many experiments such as the excess perihelion
precession of Mercury, gravitational redshift etc. at the scales
of local system, it probably needs to be modified in the
cosmological scale, especially if we want to use it to explain the
recent observational result of the acceleration. Using the idea of
Dirac cosmology with the variation of $G$ one could provide a
possible way to modify the Einstein GR.

As is well known, the Lagrangian density of Einstein GR with a
cosmological term is
\begin{equation} \label{e2}
L_E  = \frac{1}{{16\pi G}}\sqrt { - g} (R - 2\Lambda ) =
\frac{1}{{16\pi G}}\sqrt { - g} R(1 - \alpha ),
\end{equation}
where $\alpha  = 2\Lambda /R$ is a dimensionless parameter. The
cosmological constant $\Lambda$ can be generally explained as the
background fluctuation of the cosmological vacuum, and $R$ is the
4-dimensional scalar curvature. In ordinary astrophysical problems
of local systems, the magnitude of $R$ is about $8\pi G\rho  +
4\Lambda $. Since the vacuum energy density is much smaller compared
to the density $\rho$ in local system, $\alpha$ is a very small
quantity and can be neglected. But in cosmological problems, because
$\rho$ is small and has the same order as that of the vacuum energy
density, $\alpha$ cannot be neglected. In the limit of vacuum or the
matter domination area, $\rho  \to 0$, $R \to 4 \Lambda$, $\alpha$
can attain the magnitude $1/2$. This means that $\alpha$ plays an
important role at cosmological scale. However, there is only a first
order term of $\alpha$ in the equations of Einstein cosmology.
Instead of the factor $1 - \alpha$, we argue that in a perfect
cosmological theory, the Lagrangian density of the gravitational
field could contain higher-order terms of $\alpha$. We take the
Lagrangian density as
\begin{equation} \label{e3}
L_E  = \frac{1}{{16\pi G}}\sqrt { - g} Rf(\alpha ),
\end{equation}
where
\begin{equation} \label{e4}
f(\alpha ) = 1 - \alpha  - b_2 \alpha ^2  - ...
\end{equation}
The Einstein action becomes
\begin{equation} \label{e5}
S = \frac{1}{{16\pi G}}\int {d^4 x} \sqrt { - g} R(1 - \alpha  -
b_2 \alpha ^2  - ...) + \int {d^4 x} \sqrt { - g} L_M.
\end{equation}

Starting from Eq.(\ref{e5}), we can establish a Dirac cosmology
with varying $G$ and matter creation. Obviously, this theory can
unify GR and Dirac cosmology because in a local system with large
scale, $\alpha \to 0$ and the theory reduces to Einstein GR. Only
at cosmological scale, the terms $\alpha$, $\alpha ^2$... become
important and our theory reduces to Dirac cosmology.

Recently, many authors have considered the terms $R ^m(m>1)$ and/or
$R^n (n<0)$ on gravity \cite{s14}-\cite{s19}. But none of them has
connected with Dirac cosmology and the large number hypothesis. In
this paper, after introducing the higher order terms of $\alpha$, we
will establish a unified theory of GR and Dirac cosmology and
explain the acceleration of the contemporary universe, variation of
$G$, cosmological supernova Type Ia data, and the creation of
matter. We find that the addition creation of matter looks like the
dark energy and the multiplication creation of matter like the dark
matter.

The organization of this paper is as follows: we will present our
model in Sec.II. In Sec.III, by using the present cosmological
parameters as our initial conditions, we will calculate
numerically the physical quantities of our universe and study the
addition creation and multiplication creation. We will prove that
our results are in good agreement with present knowledge in
cosmology. Finally, a summary and conclusion will be given in
Sec.IV.

\section{Unified Theory of General Relativity and Dirac Cosmology}
Up to the second order of magnitude $O(\alpha ^2)$, the action of
our model is
\begin{equation} \label{e6}
S = \frac{1}{{16\pi G}}\int {d^4 x} \sqrt { - g} R(1 - \alpha  -
b_2 \alpha ^2  ) + \int {d^4 x} \sqrt { - g} L_M,
\end{equation}
where $b_2$ is a expansion parameter and we will adjust it from
supernova data later. The field equation for the metric is then
\begin{equation} \label{e7}
(1 + b_2 \alpha ^2 )R_{\mu \nu }  - \frac{1}{2}R(1 - \alpha  - b_2
\alpha ^2 )g_{\mu \nu }  + b_2 \alpha ^2 R^2 (g_{\mu \nu } \nabla
^\sigma  \nabla _\sigma   - \nabla _\mu  \nabla _\nu  )R^{ - 2}  =
8\pi GT^M _{\mu \nu },
\end{equation}
where $T^M _{\mu \nu}$ is the energy-momentum tensor of matter and
$\nabla _\mu (\mu = 0,1,2,3)$ is the covariant derivative induced by
the metric. When $ b_2 \ne 0$, the constant curvature vacuum
solutions require
\begin{equation} \label{e8}
\alpha  = \frac{{ - 1 \pm \sqrt {1 + 3b_2 } }}{{3b_2 }}.
\end{equation}
We find the constant-curvature vacuum solutions are not Minkovski
space, but rather de Sitter space ($+$) or anti-de Sitter space
($-$).

Eq.(\ref{e7}) can be rewritten as
\begin{equation} \label{e9}
G_{\mu \nu }  = 8\pi \bar G(T^M _{\mu \nu }  + \theta _{\mu \nu }
),
\end{equation}
where $G_{\mu \nu}$ is the Einstein tensor, and
\begin{equation} \label{e10}
\bar G = G/(1 + b_2 \alpha ^2 )
\end{equation}
\begin{equation} \label{e11}
\theta _{\mu \nu }  =  - \frac{\Lambda }{{8\pi G}}g_{\mu \nu }  -
\frac{1}{{8\pi G}}\sqrt {\frac{{4b_2 \Lambda ^2 (G - \bar
G)}}{{\bar G}}} g_{\mu \nu }  - \frac{1}{{8\pi }}(g_{\mu \nu }
\nabla ^\sigma  \nabla _\sigma   - \nabla _\mu  \nabla _\nu
)\frac{1}{{\bar G}}.
\end{equation}
The physical meaning of $\bar G$ and $\theta_{\mu \nu}$ is obvious.
We see from Eq.(\ref{e9}) that $\bar G$ plays the role of
gravitational constant and $\theta_{\mu \nu}$ the creation of
energy-momentum in the universe. They depend on the second order
term $b_2 \alpha^2$ of the expansion $f(\alpha)$. If $b_2=0$, $\bar
G= G=const.$, $\theta_{\mu \nu}$ reduces to the ordinary vacuum
energy cosmological term. Since $\bar G$, $\theta_{\mu \nu}$ depend
on $R$ and then are functions of time $t$, this is just the
character predicted by Dirac large number hypothesis. Futhermore,
using Eq.(\ref{e9}) and the Bianchi identity, we get
\begin{equation} \label{e12}
\nabla ^\nu  \theta _{\mu \nu }  =  - (T^M _{\mu \nu }  + \theta
_{\mu \nu } )\nabla ^\nu  \ln \bar G.
\end{equation}
$\theta_{\mu \nu}$ is no longer conserved. To understand the
physical meaning of Eq.(\ref{e12}), we rewrite it as following:
\begin{equation} \label{e12.1}
\nabla ^\nu  \theta ^{(1)} _{\mu \nu }  =  - T^M _{\mu \nu }
\nabla ^\nu  \ln \bar G
\end{equation}
\begin{equation} \label{e12.2}
\nabla ^\nu  \theta ^{(2)} _{\mu \nu }  =  - \theta _{\mu \nu }
\nabla ^\nu  \ln \bar G,
\end{equation}
where $\theta ^{(1)} _{\mu \nu}+\theta ^{(2)} _{\mu \nu}=\theta
_{\mu \nu}$. It is clear that $\theta ^{(1)} _{\mu \nu}$ is
created around the normal matter, while the creation of $\theta
^{(2)} _{\mu \nu}$ spreads over the whole universe. In the words
of Dirac Cosmology, $\theta ^{(1)} _{\mu \nu}$ is the
multiplication creation having the basic character of dark matter
and $\theta ^{(2)} _{\mu \nu}$ is the addition creation having the
basic character of dark energy. Both of them depend on the
variation of $\bar G$. In Sec.III, the suggestions that addition
and multiplication creations being dark energy and dark matter
respectively will be elaborated numerically.

\begin{figure}
\rotatebox{-90}{\resizebox{0.5\linewidth}{!}{\includegraphics*{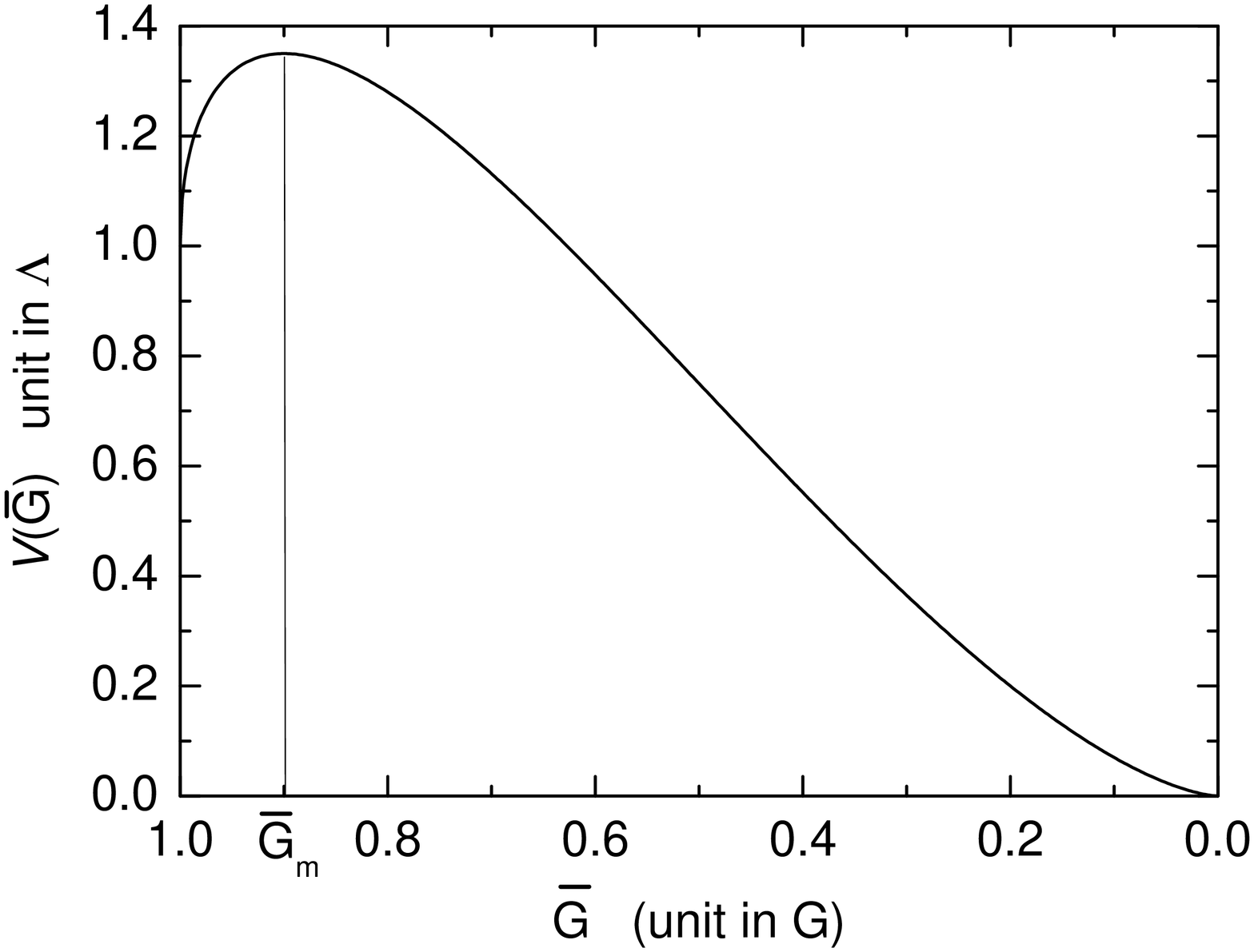}}}
\caption{The potential $V(\bar G)$ with $b_2 =1$}
\label{f1}
\end{figure}

Employing flat Robertson-Walker metric
\begin{equation} \label{e13}
ds^2  =  - dt^2  + a^2 (t)d\vec x^2 \quad (k = 0)
\end{equation}
and the perfect-fluid energy momentum tensor
\begin{equation} \label{e14}
T^M _{\mu \nu }  = (\rho _M  + p_M )U_\mu  U_\nu   + p_M g_{\mu
\nu },
\end{equation}
where $U^{\mu}$ is the 4-velocity in fluid-rest frame, $\rho_M$,
$p_M$ are the energy density and pressure respectively. Taking $p_M
= w \rho_M$ as the equation of state, $w=0$ for matter, $w=1/3$ for
radiation, we find that the field equations become
\begin{equation} \label{e15}
3H^2  = 8\pi \bar G(\rho _M  + \rho _\theta  ) = 8\pi \bar G\rho
_M  + \frac{G}{{\bar G}}V(\bar G) + 3H\frac{{d\ln \bar G}}{{dt}}
\end{equation}
\begin{equation} \label{e16}
2\dot H + 3H^2  =  - 8\pi \bar G(p_M  + p_\theta  ) =  - 8\pi \bar
Gp_M  + \frac{G}{{\bar G}}V(\bar G) - 2\frac{{\mathop {\bar
G}\limits^ \cdot  }}{{\bar G}}(\frac{{\mathop {\bar G}\limits^
\cdot  }}{{\bar G}} - H) + \frac{{\ddot {\bar G}}}{{\bar G}},
\end{equation}
where $H= \dot a/ a$, $\rho_{\theta}=\theta_{00}$,
$p_{\theta}=\theta_{ii} / g_{ii}$ $(i=1,2,3)$ are the energy
density and pressure of the created matter, respectively.
Comparing to the kinetic term $(\frac{{\dot {\bar G}}}{{\bar
G}})^2$,
\begin{equation} \label{e17}
V(\bar G) = (\frac{{\bar G}}{G})^2 \{ \Lambda  + \sqrt {4b_2
\Lambda ^2 \frac{{G - \bar G}}{{\bar G}}} \}
\end{equation}
which corresponds to a potential. Here we only consider the case
of positive curvature scalar $R$. It is easy to show that $R$ is
always positive in the evolution of the universe with matter and
radiation. The kinetic term and the potential determine the
evolution of $\bar G$ explicitly. $V(\bar G)$ will reach its
maximum at $ \bar G_m \equiv G/[1 + b_2 /(1 + \sqrt {1 + 3b_2 }
)^2 ]$ as shown in Fig.(\ref{f1}).

We also define $\rho_{\theta ^{(1)}}=\theta ^{(1)}_{00}$,
$\rho_{\theta ^{(2)}}=\theta ^{(2)}_{00}$, $p_{\theta
^{(1)}}=\theta ^{(1)}_{ii}/ g_{ii}$ and $p_{\theta ^{(2)}}=\theta
^{(2)}_{ii}/ g_{ii}$ $(i=1,2,3)$ as the energy density and
pressure of both of the multiplication and addition creation
matter. The components of $\mu=0$ of Eq.(\ref{e12.1}) and
Eq.(\ref{e12.2}) can be written as
\begin{equation} \label{e12.3}
\dot \rho _{\theta ^{(1)} }  + 3H(\rho _{\theta ^{(1)} }  +
p_{\theta ^{(1)} } ) =  - \rho _M \frac{{\mathop {\bar G}\limits^
\cdot  }}{{\bar G}}
\end{equation}
\begin{equation} \label{e12.4}
\dot \rho _{\theta ^{(2)} }  + 3H(\rho _{\theta ^{(2)} }  +
p_{\theta ^{(2)} } ) =  - \rho _\theta  \frac{{\mathop {\bar
G}\limits^ \cdot  }}{{\bar G}}
\end{equation}
The other components of Eq.(\ref{e12.1}) and
Eq.(\ref{e12.2})$(\mu=1,2,3)$ give the identities about pressures
$p_{\theta ^{(1)}}$ and $p_{\theta ^{(2)}}$ and no further
information. Since the multiplication creation $\theta ^{(1)}$
looks like dark matter, we set
\begin{equation} \label{e12.5}
p_{\theta ^{(1)}}=0 \quad p_{\theta ^{(2)}} =p_{\theta}
\end{equation}

Eq.(\ref{e15})-Eq.(\ref{e12.5}) and the equation of scalar
curvature
\begin{equation} \label{e18}
R = 6[\frac{{\ddot a}}{a} + (\frac{{\dot a}}{a})^2 ] = 6(\dot H +
2H^2 )
\end{equation}
given by the flat Robertson-Walker metric form a complete set. In
principle, we can solve the set of equations to find the behavior
of the cosmological evolution.

\section{The Numerical Solutions of the Universe With Matter And Radiation}
We now study the behavior of our universe containing matter and
radiation. To find the cosmological solution, we need the initial
conditions of Eq.(\ref{e15})-Eq.(\ref{e12.5}) in the early
universe which, unfortunately, is little known now. Hence, we have
done our numerical computation by using the present cosmological
parameters as our initial conditions, which include the Hubble
parameters constant $H_0 = 0.7 \times 100 km \cdot s^{-1}
Mpc^{-1}$, the present deceleration $q_0 = -0.5$, the cosmological
constant $\Lambda=1$ (as the scale), the density of baryon
$\Omega_b=0.05$, the ratio of the baryon-to-matter (including
baryon and multiplication creation) $\Omega_b / \Omega_m = 0.17$
and the ratio of baryon-to-photon $\eta = 6.1 \times 10
^{-10}$\cite{s22}. We will start from the present epoch and trace
back the history of our universe.

\begin{figure}
\rotatebox{-90}{\resizebox{0.5\linewidth}{!}{\includegraphics*{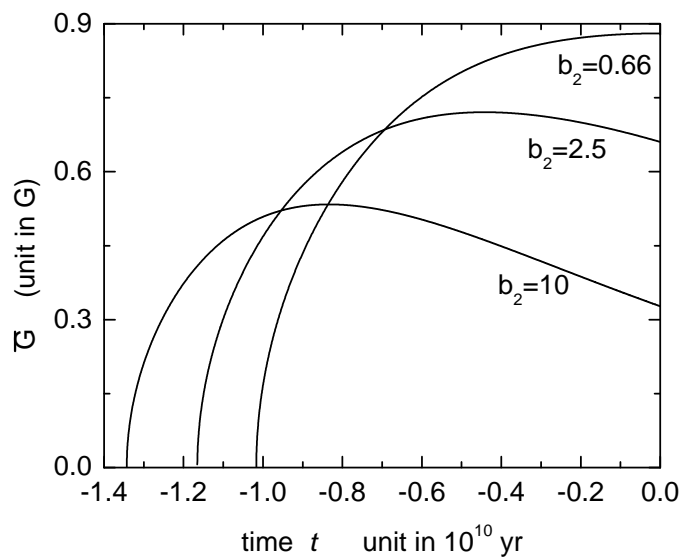}}}
\caption{$\bar G$ versus time $t$. The present time is set as $t_0
=0$. The three curves from the top to the bottom are for $b_2 =
0.66, 2.5$ and $10$.}
\label{f4}
\end{figure}
\begin{figure}
\rotatebox{-90}{\resizebox{0.5\linewidth}{!}{\includegraphics*{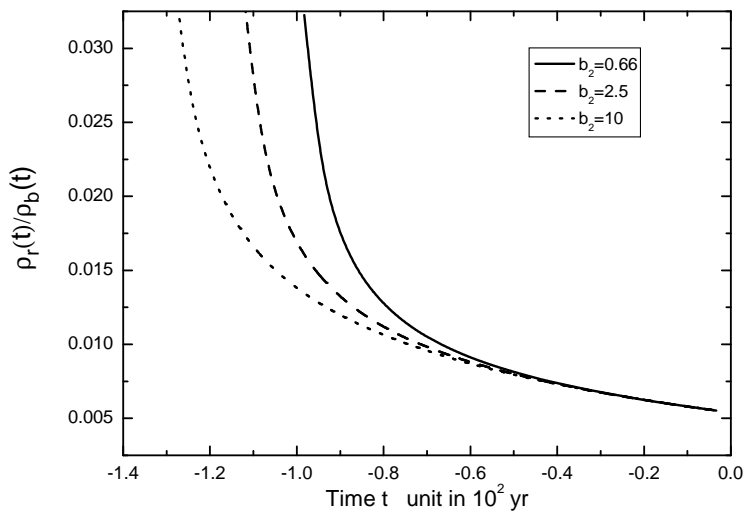}}}
\caption{The ratio of the density of radiation-to-baryon $\rho_r /
\rho_b$ versus time $t$.}
\label{f5}
\end{figure}

\begin{figure}
\rotatebox{-90}{\resizebox{0.5\linewidth}{!}{\includegraphics*{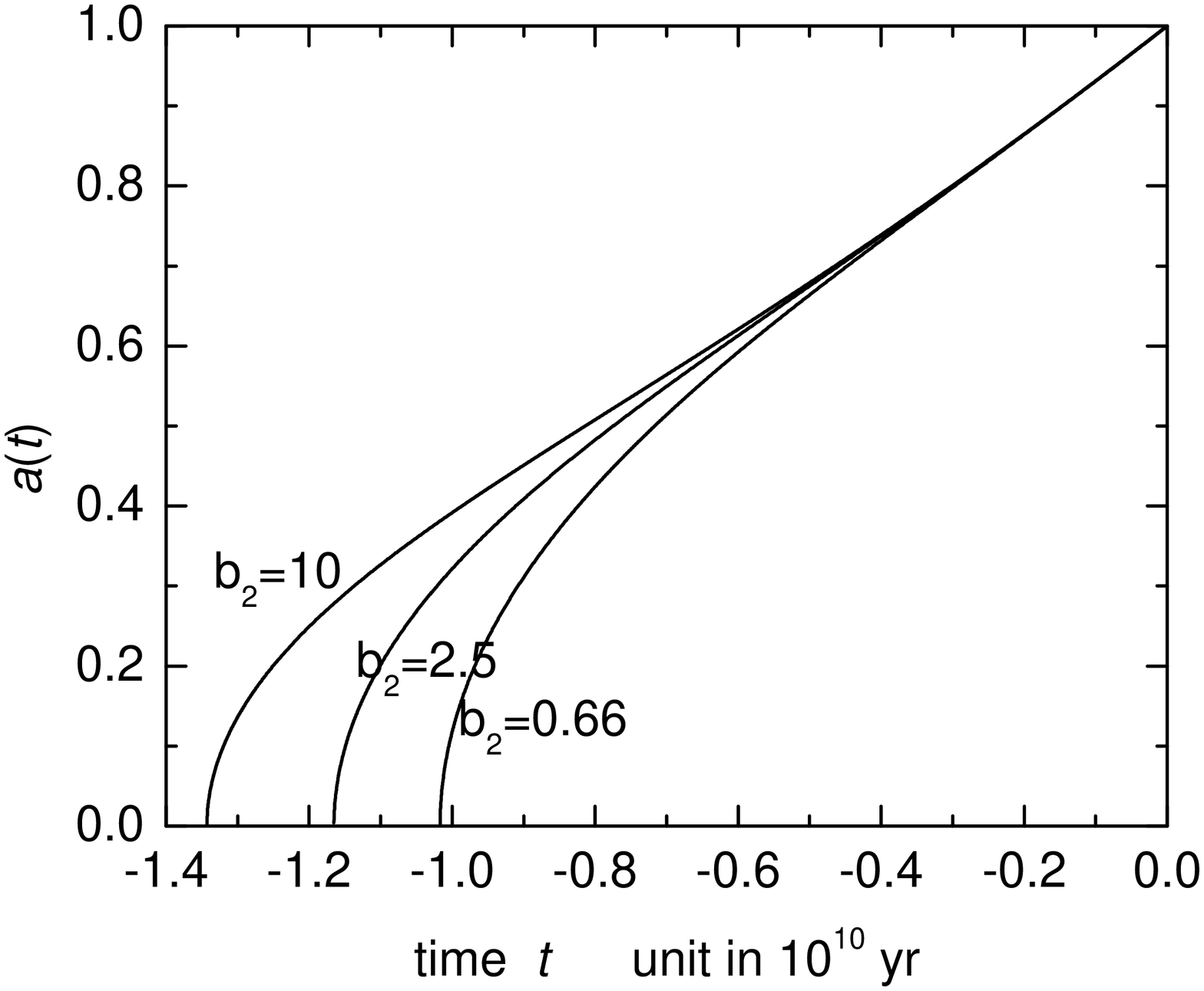}}}
\caption{The scale factor $a(t)$ versus time $t$. The three curves
from right to left correspond to $b_2=0.66,2.5$ and $10$. All of
them are asymptotic to $t^{1/2}$ as $t$ approaches the age of the
universe.}
\label{f6}
\end{figure}

\begin{figure}
\rotatebox{-90}{\resizebox{0.5\linewidth}{!}{\includegraphics*{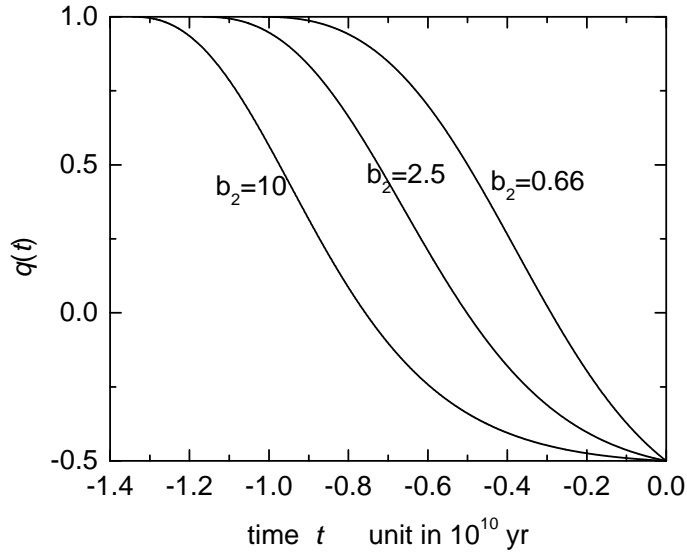}}}
\caption{The evolution of the deceleration $q(t)$. The three
curves from right to left correspond to $b_2=0.66,2.5$ and $10$.}
\label{f7}
\end{figure}

Fig.\ref{f4} shows the variation of $\bar G$ in the evolution. The
present time is set as $t_0=0$. The three curves correspond to
$b_2 = 0.66,2.5$ and $10$ respectively. We see that the
'gravitational constant' $\bar G$ increases at first when the
universe expands and then decreases at the late time. The ratio of
the density of radiation to baryon floats up quickly and the scale
factor $a(t) \to t^{1/2}$ as $t$ approaches the age of the
universe, which are shown in Fig.\ref{f5} and Fig.\ref{f6}
respectively. Therefore, the radiation dominates the early
universe. And the age of the universe given by $a(t)=0$ is $1.02
Gyr$, $1.16 Gyr$ and $1.34 Gyr$ for $b_2 = 0.66,2.5$ and $10$
respectively. Since there is no mechanism of inflation in our
model, it is easy to understand that our model can consider the
universe back to the epoch of radiation domination. Fig.\ref{f7}
shows the evolution of the deceleration factor $q$. We see that
the universe decelerates in the early era and gradually stops
decelerating and starts accelerating. This is in a good agreement
with what we understand the universe nowadays. Through the
deceleration factor $q$, we notice that the scalar curvature $R$
is always positive.


\begin{figure}
\rotatebox{-90}{\resizebox{0.5\linewidth}{!}{\includegraphics*{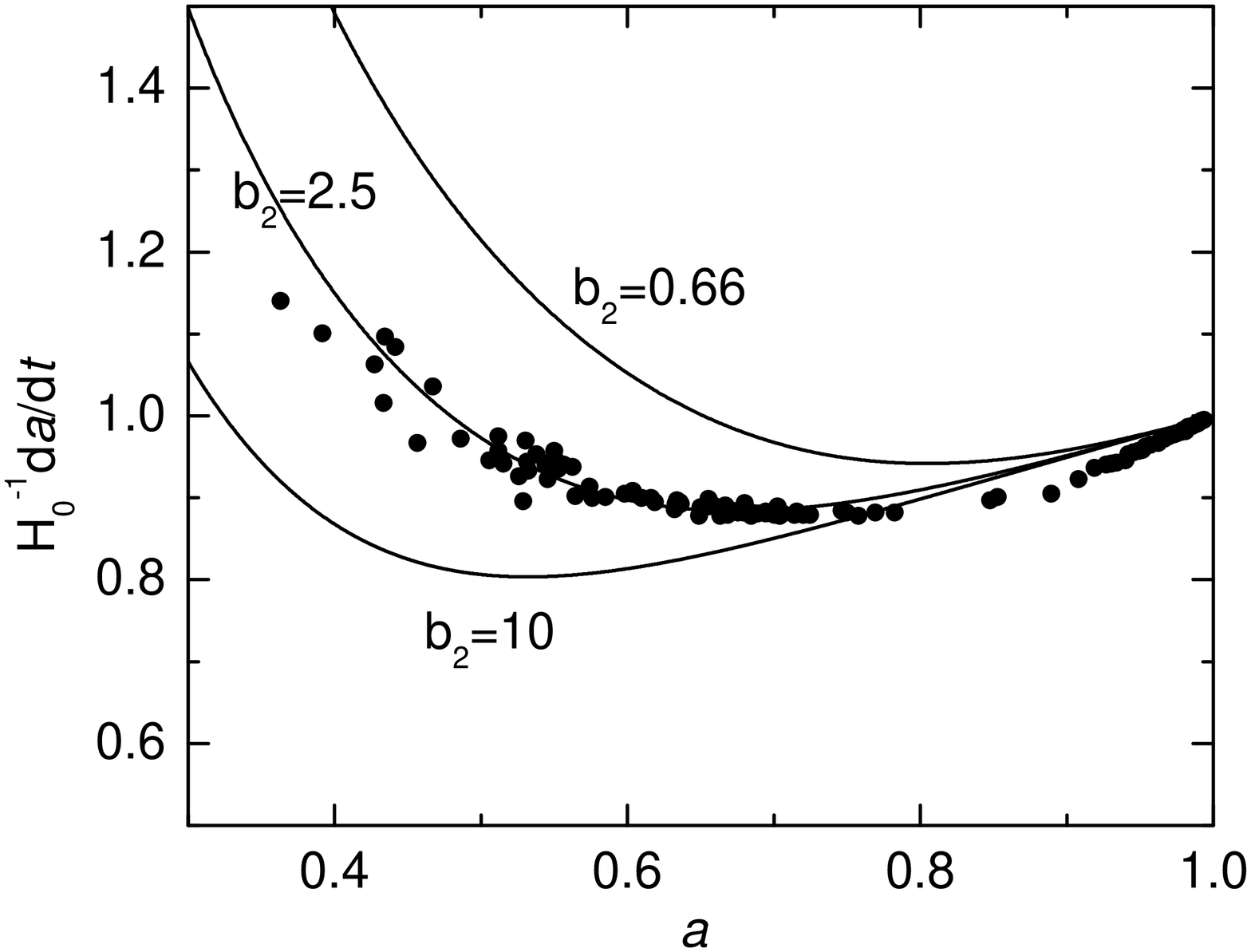}}}
\caption{The "velocity" of the scale factor $\dot a$ versus the
"position" $a$. The dots are the supernova data from
\cite{s20}\cite{s21}. The three curves from top to bottom
correspond to $b_2=0.66,2.5$ and $10$.}
\label{f8}
\end{figure}

Obviously, $b_2$ is the parameter in our model which needs to be
adjusted. To determine this parameter, let us compare our model
with the supernova data. Noting that the uncertainty of the
present value of the deceleration factor $q_0$ is still large, we
will give a better estimation of the parameter $b_2$ and $q_0$ to
fit the supernova data. Fig.\ref{f8} presents the supernova data
as a phase portrait of the universe \cite{s20}\cite{s21}, where
the universe was decelerating at high redshift and started
accelerating when it was about two-third of the present size. Our
model is favored  by this set of data : $b_2 = 2.5$ and $q_0 =
-0.5$. Comparing Fig.\ref{f8} and the curves given by Fig.2 of
ref\cite{s21}, we find that the fit of our curve is better than
that of the standard model.

\begin{figure}
\rotatebox{-90}{\resizebox{0.5\linewidth}{!}{\includegraphics*{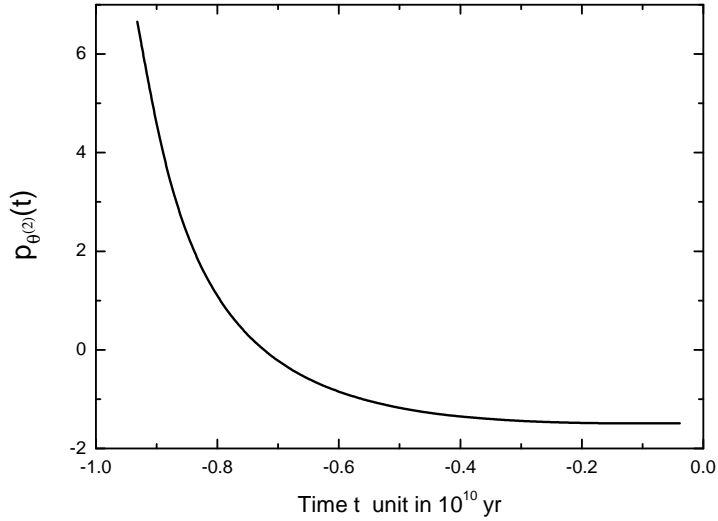}}}
\caption{The evolution of the pressure of the addition creation
$p_{\theta ^{(2)}}$ at $b_2=2.5$.}
\label{f9}
\end{figure}

\begin{figure}
\rotatebox{-90}{\resizebox{0.5\linewidth}{!}{\includegraphics*{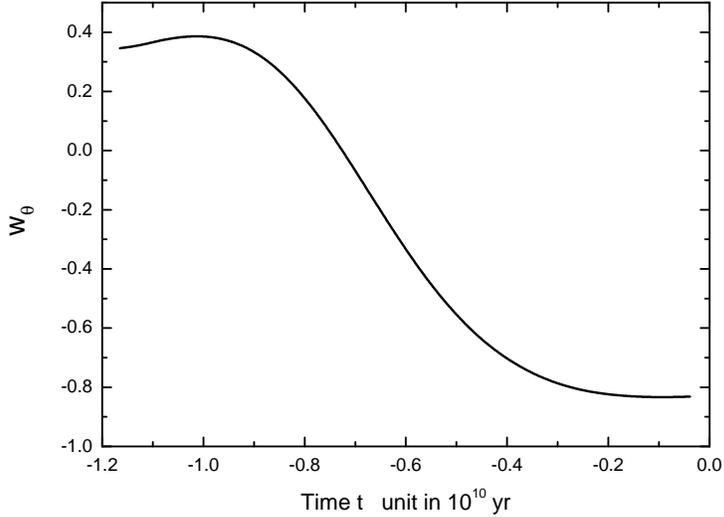}}}
\caption{The evolution of the parameter of the pressure-to-density
of the addition creation $w_ {\theta}$ versus time $t$ at
$b_2=2.5$.}
\label{f10}
\end{figure}

Fig.\ref{f9} shows the evolution of the pressure of the density of
the addition creation matter with $b_2 =2.5$. The pressure is
highly positive in the beginning of the universe, falls quickly
blow zero and has a negative value in the recent and present era.
This negative pressure looks like the character of the dark
energy. If we take $p_{\theta ^{(2)}} = w_{\theta} \rho_{\theta
^{(2)}}$ as the equation of state of addition creation, we show
the parameter of the pressure-to-density of the addition creation
$w_ {\theta}$ versus time $t$ in Fig.\ref{f10} and find
$w_{\theta}(t=0) = -0.83$ which is much smaller than $-1/3$ and is
consistent with the requirement of dark energy to explain the
accelerating universe. To compare our results with other models,
we have also calculated the ratio of the density of addition
creation and find
\begin{equation} \label{e24}
\frac{{\rho _{\theta ^{(2)} } }}{{\rho _{\theta ^{(1)} } + \rho
_{\theta ^{(2)} }  + \rho _r  + \rho _b }} = \Omega_ {\Lambda} =
0.8
\end{equation}
and that of the density of matter (multiplication creation and
normal matter) in our model reads
\begin{equation} \label{e25}
\frac{{\rho _{\theta ^{(1)} }  + \rho _r + \rho _b }}{{\rho
_{\theta ^{(1)} } + \rho _{\theta ^{(2)} }  + \rho _r  + \rho _b
}} = \Omega_ M = 0.2.
\end{equation}
These results have the same magnitude as that dark energy and dark
matter respectively. Hence, we suggest that the dark energy comes
from the addition creation and the dark matter from the
multiplication creation.

\section{Summary And Discussion}
We have suggested a model to unify the Einstein GR and Dirac
Cosmology. In local system, our theory reduces to GR, but in the
cosmological scale, our theory refers to the Dirac cosmology. The
variation of the gravitational constant comes from the scalar
curvature. In the local system, the variation of the gravitational
constant is negligible. This result is in good agreement with
present experiments. But in the cosmological scale, the change of
$\bar G$ is remarkable. The acceleration of the present universe
is a proof of the decrease of the gravitational constant $\bar G$,
because the decrease of $\bar G$ corresponds to an effective
repulsion.

After introducing the term $-b_2 \alpha^2$ in our theory and fit the
adjust parameter $b_2$ by supernova data, we have calculated the
physical quantities of $\bar G(t)$, $a(t)$, $q(t)$, $\rho_r(t)/
\rho_b(t)$ by using the data of present epoch as the initial
conditions. We have found that the results are self-consistent and
in good agreement with present knowledge of cosmology.

According to Dirac large number hypothesis, matter will be created
in the universe. We have calculated the matter comes from addition
creation and that from multiplication. An interesting picture in our
theory is that the addition creation, which spreads over the
universe, looks like the dark energy and the multiplication
creation, which clusters around the normal matter, like the dark
matter. We have found that the pressure of addition creation has a
big positive value initially and fall down quickly to become a
negative value. It has a negative pressure region corresponding $ 0
> t >  - 0.74 \times 10^{10} yr$. The equation of state parameter
$w_{\theta}=-0.83$ at present. This value is in agreement with
present dark energy models. We have also calculated the ratio of the
density of the addition creation and found $\Omega_{\Lambda}=0.8$.
The same parameter but for multiplication creation, radiation and
normal matter has also been computed, which reads $\Omega_m=0.2$.
Both of them have the same magnitude of the observational value of
the dark energy and matter. This result suggests that the dark
energy and dark matter are just the addition creation and the
multiplication creation in Dirac cosmology.

Finally, we would like to emphasize that this model cannot be
extended to the big bang epoch. We have not add the terms with $R^m
(m>1)$ (or $\alpha ^{-n}(n \ge 1)$ ) in the Lagrangian density.
Obviously, these terms are very important to the very early
universe, especially in the inflationary epoch. Our model can only
be used in the time evolution regions starting from the radiation
dominated epoch to the present time.

\begin{acknowledgments}

This work was supported in part by NNSF of China, by the National
Basic Research Program 2003CB716300 and the Foundation of
Education of Ministry of China. B. Wang's work was also supported
in part by Shanghai Education Commission.

\end{acknowledgments}


\end{document}